\documentclass[12pt]{article}
\usepackage{epsfig}
\usepackage{float}
\usepackage{graphicx}
\usepackage{cite}
\begin{document}
\rightline{NKU-2021-SF1}
\bigskip

\newcommand{\be}{\begin{equation}}
\newcommand{\en}{\end{equation}}
\newcommand{\bea}{\begin{eqnarray}}
\newcommand{\ena}{\end{eqnarray}}
\newcommand{\noi}{\noindent}
\newcommand{\ra}{\rightarrow}
\newcommand{\bib}{\bibitem}
\newcommand{\refb}[1]{(\ref{#1})}
\newcommand{\bff}{\begin{figure}}
\newcommand{\eff}{\end{figure}}

\begin{center}
{\Large \bf Thermodynamics and Heat Engines of Black Holes with Born-Infeld-type Electrodynamics}
\end{center}
\hspace{0.4 cm}
\begin{center}
Leonardo Balart\footnote{leonardo.balart@ufrontera.cl}\\
{\small \it Departamento de Ciencias F\'{\i}sicas, \\
Facultad de Ingenier\'{\i}a y Ciencias \\ \small Universidad de La Frontera, Casilla 54-D \\
Temuco, Chile.}\\
\vspace{0.3 cm}
Sharmanthie Fernando\footnote{fernando@nku.edu}\\
{\small\it Department of Physics, Geology  \& Engineering Technology}\\
{\small\it Northern Kentucky University}\\
{\small\it Highland Heights, Kentucky 41099, U.S.A.}\\
\end{center}

\begin{center}
{\bf Abstract}
\end{center}

In this paper we have studied electrically charged black holes  in a new model of nonlinear electrodynamics  introduced by Kruglov in Ref.~\cite{krug1}. There are two parameters for the theory and the black hole could have  up to two horizons. Thermodynamics is studied in the extended phase space where the pressure is proportional to the cosmological constant. First law and the Smarr formula are derived. There are phase transitions similar to the Van der Waals liquid-gas phase transitions. Black hole is also studied as a heat engine and have discussed how the parameters in the nonlinear electrodynamics theory effect the efficiency of  the heat engine.

\hspace{0.7cm}



\section{Introduction}
\label{intro}

The Born-Infeld theory~\cite{born} is a model of non-linear electrodynamics which was proposed to avoid singularities of the Maxwell model. This theory is characterized by a non-linear parameter  $b$ which represents the value of the electromagnetic field strength at the origin of a point source. In the limit $b \rightarrow \infty$ the Maxwell electrodynamics is recovered.
This theory has received attention from different contexts~\cite{Hoffmann:1935ty,Salazar:1987ap,Cataldo:1999wr,GarciaSalcedo:2000eb,Fernando:2003tz,Miskovic:2008ck,Tseytlin:1997csa,Cecotti:1986gb,Callan:1997kz}. In recent times alternative theories of nonlinear electrodynamics have been presented: some of them are also characterized by a nonlinear  parameter~\cite{Soleng:1995kn,Hendi:2012zz,gaete,Kruglov:2014iqa,Kruglov:2015fcd,Kruglov:2016ezw,Kruglov:2018lct}.
Other models have emerged in the context of regular black holes~\cite{AyonBeato:1998ub,AyonBeato:1999rg,Bronnikov:2000vy,Dymnikova:2004zc,Balart:2014jia,Balart:2014cga}. However,  as shown in Ref.~\cite{Balart:2017dzt} they cannot include a nonlinearity parameter
that allows to obtain the Reissner-Nordstr\"om solution in the limit where it tends to infinity, as in the case of the nonlinearity parameter $b$ of the Born-Infeld electrodynamics, or as in the black hole solution with nonlinear electrodynamics given in Ref.~\cite{Gullu:2020qni}, where the Reissner-Nordstr\"om case is recovered when the nonlinearity parameter tends to zero.


Recently a black hole coupled to a Born-Infeld type electrodynamic model has been proposed in Ref.~\cite{krug2}  that, in addition to the parameter $\beta$ which is related to the maximal electromagnetic field strength, depends on another extra parameter, denoted by $\sigma$, and where the Born-Infeld black hole corresponds to one of the values of $\sigma$. This new family of solutions has two characteristics that make it interesting in order to deepen its study. On the one hand, depending on the values of the extra parameter, it is possible to have solutions of an electric field that can be amplified (where therefore the regularity at the origin is lost) or screened with respect to the electric field of the Reissner-Nordstr\"om solution. This is something that deserves attention, for a better understanding about vacuum polarization in the context of black holes itself~\cite{mann4,Ruffini:2013hia}, as well as its relationship with the deviation of the electric field with respect to the Coulomb field. On the other hand, it is also worth analyzing the model from the point of view of the energy conditions that the respective energy-momentum tensor satisfies~\cite{Hawking:1973uf}. In particular, fulfilment of the dominant energy condition depends on the values that the parameter $\sigma$ takes. In respect of the latter, the relationship that the screening or the amplification of the electric field may have with the energy conditions that the corresponding energy-momentum tensor fulfills can be studied. More details on this will be presented in a later work.


One of the goals in this paper is to study the electrically charged black hole in this new electrodynamics in the extended phase space with $ P = - \frac{\Lambda}{ 8 \pi} $. In the extended phase space, first law of black hole thermodynamics has an extra tern corresponding to $ V dP$. Here $M = U + P V$ is the enthalpy and not the internal energy $U$. Many black holes have demonstrated phase transitions and critical behavior between large and small black holes in this context. Such phase transitions are similar to van der Waals liquid-gas phase transitions. There are large number of work related to this topic: we would only give few of such work as references as~\cite{mann4,mann,mann5,hen1,cao,liu2,hendi,mo,zhang,li,cai,azg,pou}. There is a review by Kubiznak et.al.~\cite{kub} which gives more references and a comprehensive review on the topic.


It should be mentioned that several works have considered AdS black holes as heat engines~\cite{johnson1,johnson2,johnson4,Johnson:2016pfa,yerra,sun,mo5,meng,meng2,mann2,Balart:2019uok}. Here the central idea is that if the extended phase space is considered one can extract mechanical work from a AdS black hole when one consider it as a  heat machine. To compute the efficiency of the heat engine one must define a thermodynamic cycle in phase space. In this paper we will present a  Born-Infeld-type black hole given by Kruglov~\cite{krug1} as a heat engine choosing one of the cycles proposed by Johnson 
in Ref.~\cite{johnson1}. This cycle is composed of two isobaric and two isochoric paths.


The paper is organized as follows: in section 2, the black hole considered is presented. In section 3 thermodynamics of the black holes are given and in section 4 P-V criticality is discussed. In section 5 the black holes are studied in the context of heat engines and finally in section 6 the conclusion is given.


\section{Black holes with Born–Infeld-type electrodynamics}
In general the action of the 3+1-dimensional Einstein theory coupled with a nonlinear electrodynamics model is given by
\begin{equation} 
S = \int d^4x \sqrt{-g} \left[ \frac{(R - 2 \Lambda)}{16 \pi G} + \mathcal{L(F)} \right] 
\end{equation}
Here, $g$ is the determinant of the metric tensor and $\Lambda = -\frac{3}{l^2}$ is the cosmological constant and $\mathcal{L(F)}$ is the the Lagrangian depending on the invariant $\mathcal{F} = F_{ \mu \nu} F^{ \mu \nu}$.
The variation of the action with respect to the gravitational field yields the Einstein equations
\begin{equation} 
G_{\mu\nu} + \Lambda g_{\mu\nu} = 8 \pi T_{\mu\nu}
\label{einstein-eq-1}
\end{equation}
\begin{equation} 
T_{\mu\nu} = g_{\mu\nu}  \mathcal{L(F)} - F_{\mu\rho} F_\nu^{\,\, \rho} \mathcal{L},_\mathcal{F}
\label{einstein-eq-2} \, .
\end{equation}
And the variation with respect to the electromagnetic field gives
\begin{equation} 
\nabla_\mu (F^{\mu\nu} \mathcal{L},_\mathcal{F}) = 0
\label{electrom-eq} 
\end{equation}
where $\mathcal{L},_\mathcal{F}$ is the derivative of $\mathcal{L(F)}$ with respect to $\mathcal{F}$.

In the present study we will restrict ourselves to the nonlinear electrodynamic model described by the Lagrangian density
\begin{equation} 
\mathcal{L}(\mathcal{F}) = \beta^2\left[1 - \left(1 + \frac{\mathcal{F}}{\beta^2 \sigma}\right)^\sigma \right]
\label{lagrangian-kr} \, .
\end{equation}
This Lagrangian density can be obtained from the electrodynamic model introduced by Kruglov in Ref.~\cite{krug1} and whose Lagrangian density is given as
\begin{equation}
\mathcal{L(F)} = \frac{1}{\beta} \left[  1 - \left( 1 + \frac { \beta \mathcal{F}}{ \sigma} - \frac{ \beta \gamma \mathcal{G}^2}{ 2 \sigma} \right)^{\sigma} \right]
\label{lagrangian-kr-g} 
\end{equation}
where $\mathcal{F} = F_{ \mu \nu} F^{ \mu \nu}$, $\mathcal{G} = \frac{1}{4} F_{\mu \nu} \tilde{F}^{\mu \nu}$ and $\beta, \gamma, \sigma$ are  parameters of the theory. 
In particular by choosing $\gamma = 0$ and doing the replacement $\beta \rightarrow 1 / \beta^ 2$ in Eq.~(\ref{lagrangian-kr-g}), let us obtain Eq.~(\ref{lagrangian-kr}).

When $\beta = \gamma $ and $\sigma = \frac{1}{2}$ in Eq.~(\ref{lagrangian-kr-g}), the model becomes the Born-Infeld electrodynamics. Hence the above new model is named Born-Infeld-type electrodynamics. A two parameter model was considered in Ref.~\cite{gaete} which is close to the above mentioned model. In Ref.~\cite{krug2}, Kruglov investigated a magnetic black hole arising from the above model coupled to general relativity. 
In Refs.~\cite{gaete,Akmansoy:2017vcy,Akmansoy:2018xvd,Neves:2021tbt}, the possible constraints for the parameters of this model (and also of other nonlinear electrodynamic models) are analyzed.

\begin{figure*} 
\begin{center}
\includegraphics{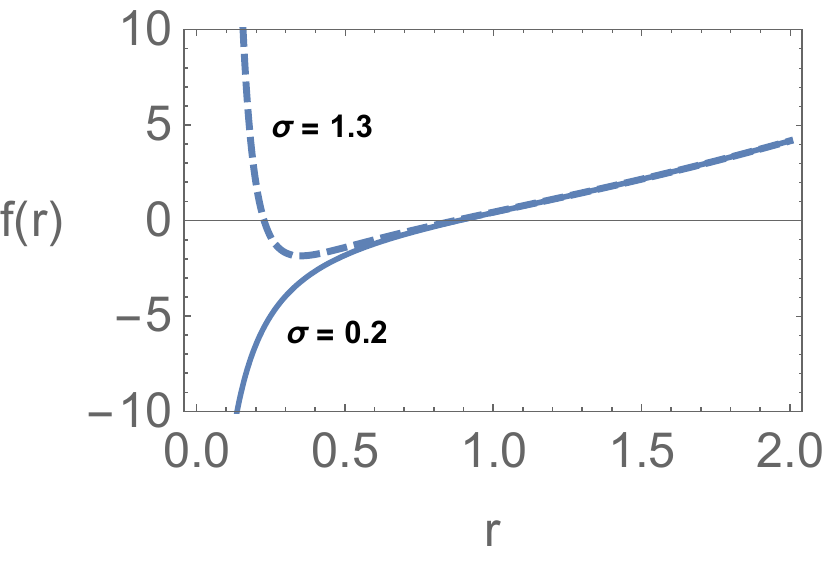}
\caption{The figure shows  $f(r)$ vs $r$. Here  $q= 0.434, M= 0.878$, and $\beta = 0.425$. The plots correspond to two values of $\sigma$ as given in the figure.}
\label{frvsr}
 \end{center}
 \end{figure*}

Static spherically symmetric black holes for the above theory is given by the metric $ds^2 = -f(r)dt^2 + f^{-1}(r)dr^2 + r^2(d\theta^2 + \sin^2 \theta d\phi^2)$. This ansatz in the Einstein equations~(\ref{einstein-eq-1}) and using the Lagrangian density~(\ref{lagrangian-kr}) allows us to obtain the metric function
\begin{eqnarray}
f(r) = 1 - \frac{2 M}{r} + \frac{r^2}{l^2} - \frac{2 \beta^2 r^2}{3} \left[ \, _{2}F_1\left(-\frac{3}{4},-\sigma;\frac{1}{4};-\frac{q^2}{2 \sigma \beta^2 r^4}\right) - 1 \right]
\,\,\label{metric-f-kr} \, 
\end{eqnarray}
where $M$ represents the ADM mass, $q$ the electric charge and $_{2}F_1(a,b;c;z)$ is the Gauss hypergeometric function.
In Fig.~$\ref{frvsr}$, the function $f(r)$ is plotted for two values of $\sigma$. For small $\sigma$, the black hole has one horizon and behaves similar to the Schwarzschild AdS black hole. For large $\sigma$, the black hole could have two horizons and behave as the Reissner-Nordstr\"om-AdS black hole.

\begin{figure} 
\begin{center}
\includegraphics{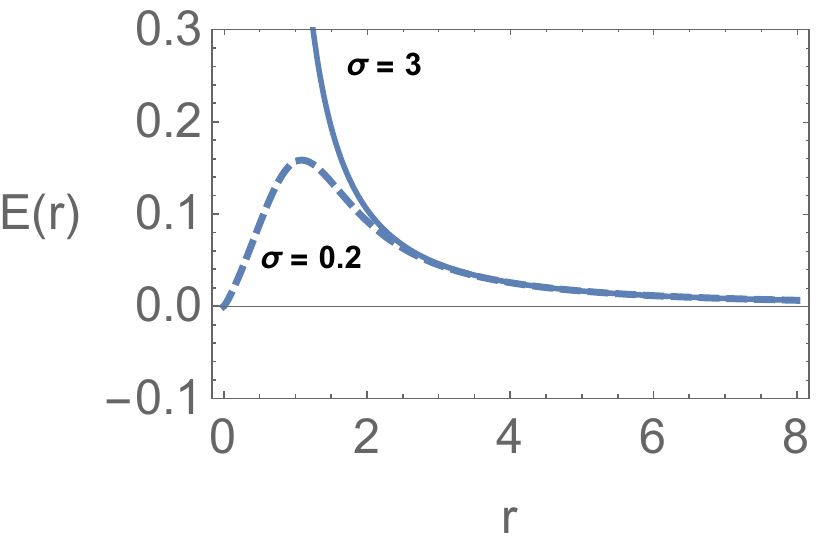}
\caption{The figure shows  $E(r)$ vs $r$. Here  $q= 0.411$ and $\beta = 0.425$. The plots corresponds to two values of $\sigma$ as given in the figure.}
\label{evsr}
 \end{center}
 \end{figure}

The electromagnetic tensor takes the form $F_{\mu \nu} = E(r) \left( \delta^t_{\mu} \delta^r_{\nu} - \delta^t_{\nu} \delta^r_{\mu} \right)$, where $E(r)$ is the electric field, that is, $\mathcal{F} = -2 E^2(r)$, which together with Eq.~(\ref{electrom-eq})  allows us to obtain the corresponding electric field  for the Born-Infeld-type electrodynamics 
\begin{equation}
E(r) = \frac{2 q r^2 \beta^2 \sigma  }{q^2+2\sigma \beta^2 r^4 }\left(1 + \frac{q^2}{2\sigma \beta^2 r^4}\right)^{\sigma}
\,\,\label{electric-field-kr} \,  .
\end{equation}
The electric field  asymptotically behaves as the one for the Reissner–Nordstr\"om-AdS black hole as,
\begin{equation}
E(r) = \frac{q}{r^2} - \frac{q^3 (1- \sigma)}{2 \sigma \beta ^2 r^6} + O(1/r^{10})
\,\,\label{electric-field-asymp-kr} \,  .
\end{equation}
In Fig.~$\ref{evsr}$, the electric field is plotted for two values of $\sigma$. For small $\sigma$, the electric field is finite at $r =0$ similar to  Born-Infeld electrodynamics. For large $\sigma$, the electric field is infinite at $r=0$ similar to Coulomb  electric field.

\section{Thermodynamics of the black hole with Born–Infeld-type electrodynamics}

In this section, we will derive thermodynamic quantities, first law and the Smarr formula for the Born-Infeld-type black hole. We will be studying thermodynamics in the extended phase space where the thermodynamic pressure is given by $P = -\Lambda/(8\pi) = 3/(8 \pi l^2)$.
The conjugate quantity to $P$ is $V = \frac{ 4 \pi r_+^3}{ 3}$.  The Hawking temperature is given by (after replacing $l$ in terms of $P$),
\begin{equation}
 T =  \frac{ 1}{ 4 \pi} \left | \frac{ d f(r)}{ dr} \right |_{r_+} = \frac{1}{4 \pi  r_+}
\left[1 + 8 \pi P r_+^2 + 2 r_+^2 \beta^2 \left(1 - \left(1 + \frac{q^2}{2 r_+^4 \beta^2 \sigma}\right)^{\sigma} \right)\right]
\,\,\label{temperature-kr} \,  
\end{equation}
where $r_+$ is the black hole horizon.  The entropy of the black hole, $S = \pi r_+^2$ is conjugate to the temperature.

The black hole horizon $r_+$ satisfies $f(r_+) = 0$. Then from Eq.~(\ref{metric-f-kr}), we can arrive at a formula for $r_+$ as a function of the parameters $l$, $q$ $\beta$, $\sigma$ and the ADM mass $M$. And from here we obtain the following expression for $M$
\begin{equation}
M = \frac{r_+}{6} \left[-2 r_+^2 \beta ^2  \, _{2}F_1\left(-\frac{3}{4},-\sigma;\frac{1}{4};-\frac{q^2}{2 \sigma \beta^2 r_+^4}\right) + 8 \pi  P r_+^2 + 2 r_+^2 \beta^2 + 3\right]
 \,\,\label{mass-BH-kr} \, 
\end{equation}
The electric potential $\Phi$ on the outer horizon, which is conjugate to the charge $q$ is computed as,
\begin{eqnarray}
\Phi(r_+) = \int_{r_+}^{\infty} E(r) dr =  \frac{\beta^2 r_+^3}{2 q} \left[\left(1 + \frac{q^2}{2 \sigma \beta^2 r_+^4}\right)^{\sigma}-\, _{2}F_1\left(-\frac{3}{4},-\sigma;\frac{1}{4};-\frac{q^2}{2 \sigma \beta^2 r_+^4}\right)\right]
\,\,\label{potential-kr} \,  
\end{eqnarray}
We have defined a new quantity $B$ as the conjugate quantity to $\beta$.  To write the first law, we will treat $M$ as the enthalpy given by $ M = U + P V$. Here, $U$ is the internal energy of the black hole. 

Rewriting the mass $M$ as
\begin{equation}
M(S, P, q, \beta) = \frac{\sqrt{S}}{6\sqrt{\pi}} \left[-\frac{2 S \beta ^2}{\pi}  \, _{2}F_1\left(-\frac{3}{4},-\sigma;\frac{1}{4};-\frac{\pi^2 q^2}{2 \sigma \beta^2 S^2}\right) + 8  P S + \frac{2 S \beta^2}{\pi} + 3\right]
 \,\,\label{mass-BH-kr-2} \, ,
\end{equation}
the differentiation allows us to obtain
\begin{equation}
dM = \left ( \frac{ \partial M} { \partial S } \right) _{q,P,\beta}  dS + \left ( \frac{ \partial M} { \partial P } \right) _{S,q, \beta} dP + \left ( \frac{ \partial M} { \partial q } \right) _{S,P, \beta}  dq
+ \left(\frac{\partial M}{\partial \beta}\right)_{S,q,P} d\beta
\,\,\label{1-law-gral} \,  ,
\end{equation}
with
\begin{equation}
\left ( \frac{ \partial M} { \partial S } \right) _{q,P,\beta} =  T  \,\,\,\,\,\,\,\,\,\,\,\,\,\,
\left ( \frac{ \partial M} { \partial q } \right) _{S,P, \beta} = \Phi  \,\,\,\,\,\,\,\,\,\,\,
\left ( \frac{ \partial M} { \partial P } \right) _{S,q, \beta} =  V \,  ,
\end{equation}
which are compatible with the expressions given in Eqs.~(\ref{temperature-kr}) and~(\ref{potential-kr}) and the volume of the black hole.
We have also defined a new quantity $B$ as the conjugate quantity to $\beta$ following Refs.~\cite{mann4,YiHuan:2010zz}. 
Thus, we can compute the variable $B$ as,
\begin{equation}
B = \left(\frac{\partial M}{\partial \beta}\right)_{S,q,P} = \frac{\beta r_+^3 }{6}  \left[4 - \, _{2}F_1\left(-\frac{3}{4},-\sigma;\frac{1}{4};-\frac{q^2}{2 \sigma \beta^2 r_+^4}\right) - 3 \left(1 + \frac{q^2}{2 \sigma \beta^2 r_+^4}\right)^{\sigma}\right]
\,\,\label{B-evaluated-kr} \,  .
\end{equation}
Consequently, these quantities satisfy the first law of thermodynamics 
\begin{equation}
dM = TdS +VdP + \Phi dq + B d\beta
\,\,\label{1-law-kr} \,  .
\end{equation} 
From Eq.~(\ref{mass-BH-kr-2}), we could obtain the Smarr formula following the scaling argument given in Ref.~\cite{Kastor:2009wy}. Carrying out the dimensional analysis, we note that the quantities obey the following scaling relations $S \propto l^2$,  $P \propto l^{-2}$, $q \propto l^1$ and $\beta \propto l^{-1}$, where $l = [\mbox{lenght}]$, therefore the mass is a homogeneous function of degree 1. Thus, the Euler theorem implies
\begin{equation}
M = (2)\left ( \frac{ \partial M} { \partial S } \right)   S + (-2) \left ( \frac{ \partial M} { \partial P } \right)  P + (1)\left ( \frac{ \partial M} { \partial q } \right)   q
+ (-1) \left(\frac{\partial M}{\partial \beta}\right) \beta
\,\,\label{smarr-scaling} \,  ,
\end{equation}
So one can write the Smarr formula~\cite{Smarr:1972kt} as  
\begin{equation}
M = 2 TS - 2VP + \Phi q - B \beta 
\,\,\label{smarr-formula} \,  .
\end{equation}

Recently, some results on thermodynamic relations have been obtained for black holes with nonlinear electrodynamics. For example, in Ref.~\cite{Nam:2018sii} the author considers the solution given in Ref.~\cite{AyonBeato:1998ub}, and uses a different thermodynamic approach than we consider, where the thermodynamic volume of the black hole is not the usual $4 \pi r_+^3/3$, thus achieving to obtain a first law without adding new quantities and according to this the Smarr formula turns out to be $M = 2TS -2VP + \Phi q$. In Ref.~\cite{Kuang:2018goo} a new black hole solution with nonlinear electrodynamics is presented, which depends on a nonlinearity parameter such that if it is zero then the Maxwell case recovers. Here, as in  Refs.~\cite{Gullu:2020qni,mann4} and our thermodynamic approach, a new term appears in the Smarr formula that is related to the quantity conjugated to the nonlinearity parameter. For its part, in Ref.~\cite{Jawad:2018cdh}, the first law and the Smarr formula are modified when a black hole with non-linear electrodynamics is considered, due to a new thermodynamic quantity, the surface tension. 
Also in the context of black hole solutions with nonlinear electrodynamics, in Refs.~\cite{Singh:2019tgw,Ma:2014qma}, the first law is modified with a quantity that multiplies $dM$. 
In Ref.~\cite{Lan:2020wpv} it is considered that the entropy for regular black holes consists of two terms, one of them the usual and the other corresponds to a deviation, and when the black hole solution is obtained by coupling a nonlinear electrodynamic model, then it is also considered a corrected term for the potential electric.
As future work one could take those thermodynamic approaches that are different from the one we are considering in the present work and make a detailed analysis for the case of black holes with Born-Infeld-type electrodynamics.


\section{ P-V criticality of the black hole}

\begin{figure} 
\begin{center}
\includegraphics[scale=0.7]{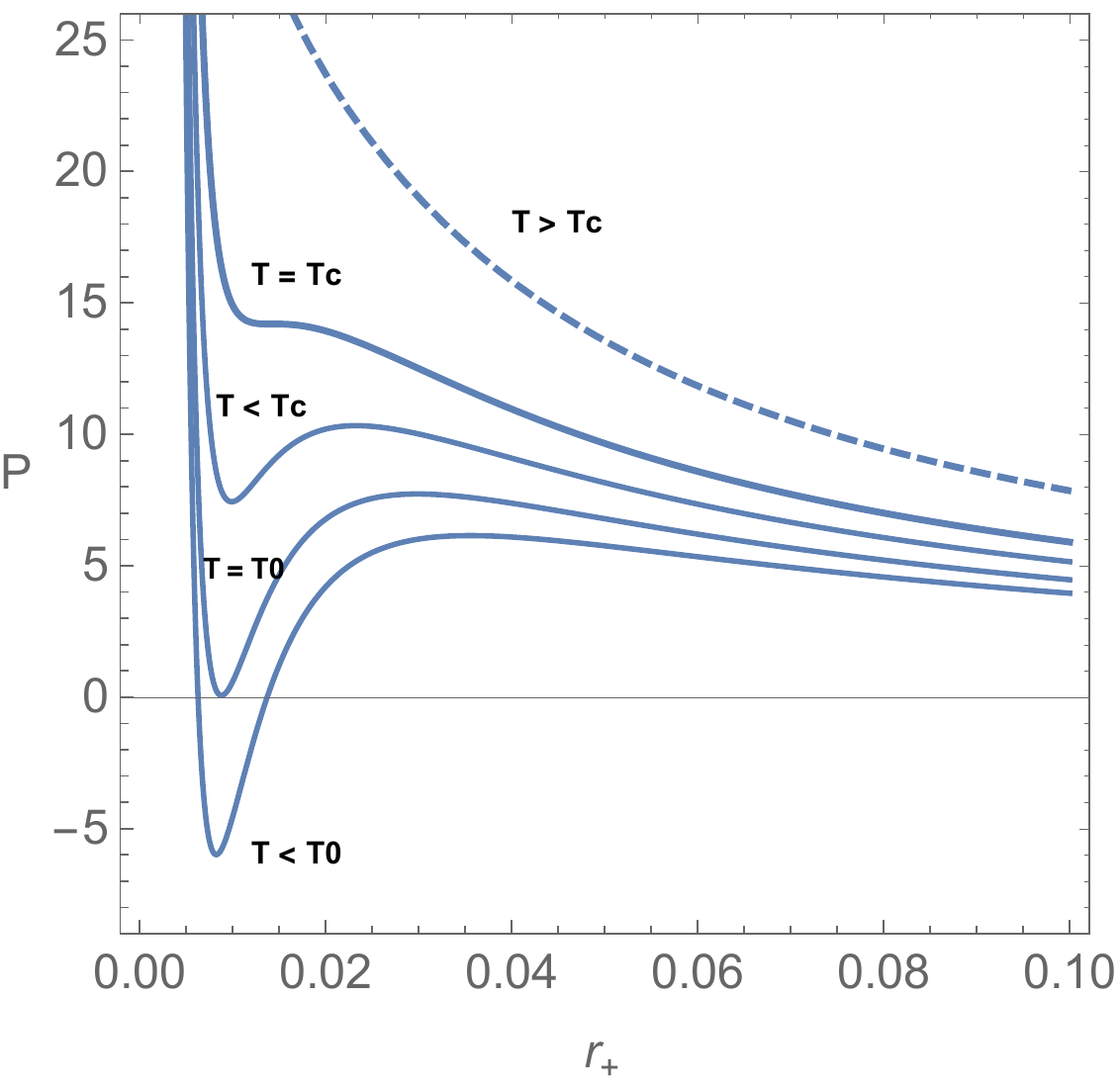}
\caption{The figure shows  $P$ vs $r_+$ for varying temperature. Here  $\sigma = 0.539, q= 0.41$, and $ \beta = 0.536$}
\label{pvsr}
 \end{center}
 \end{figure}

\begin{figure}
\begin{center}
\includegraphics[scale=0.9]{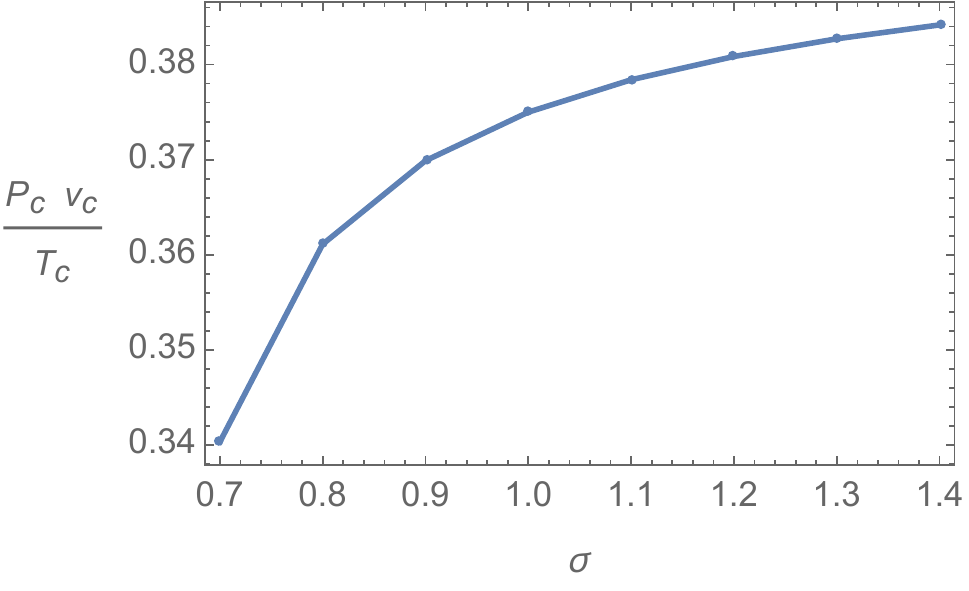}
\caption{The figure shows  $\frac{P_c v_c}{T_c}$ vs $\sigma$. Here  $q= 0.41$, and $\beta = 0.536$}
\label{ratiovssigma}
 \end{center}
 \end{figure}

In this section, we will study the behaviour of pressure. The equation of state for a Born–Infeld-type black can be described from Eq.~(\ref{temperature-kr}) as
\begin{equation} 
P = \frac{2 r^2 \beta^2 \left[\left(1 + \frac{q^2}{2 \sigma \beta^2 r_+^4}\right)^{\sigma} - 1\right] + 4 \pi  r_+ T-1}{8 \pi  r_+^2}
\,\,
\label{pressure} \,  
\end{equation}
For certain values of parameters in the theory, there are critical values for P vs $r_+$ graph as shown in Fig.~$\ref{pvsr}$.
 In order to calculate the critical values and to study  law of corresponding states, we first identify specific volume $v = 2 r_+$. Then replacing $r_+ = \frac{v}{2}$ in Eq.~(\ref{pressure}), the equation of state can be rewritten as follows:
\begin{equation}
P = \frac{ -1 + 2 \pi T v + \frac{ v^2 \beta^2}{2} \left( ( 1 + A ) ^{\sigma} -1 \right)}{ 2 \pi v^2}
\end{equation}
where $A = \frac{ 8 q^2}{ v^4 \beta^2 \sigma}$.
Since $v \propto r_+$, the graph in Fig.~$\ref{pvsr}$ will be similar in shape to a graph for $P$ vs $v$. Critical value $t_c$ occurs when the $P$ vs $v$ graph has an inflection point. At the critical point,
\begin{equation} \label{critical}
\frac{\partial P}{\partial v} = \frac{ \partial^2 P}{ \partial v^2} =0
\end{equation}
Solving the two equations obtained due to the conditions in Eq.~(\ref{critical}),  critical temperature $T_c$ and  critical pressure $p_c$ are given as follows:
\begin{equation}
T_c = \frac{ v_c^4 \beta^2 \sigma + q^2 ( 8 - 8 v_c^2 \beta^2 ( 1 + A_c)^{\sigma} \sigma )}{ \pi v_c ( 8 q^2 + v_c^4 \beta^2 \sigma)}
\end{equation}
\begin{equation}
P_c = \frac{ -2 + 4 \pi T_c v_c + v_c^2 \beta^2 ( -1 + ( 1 + A_c)^{\sigma})}{ 4 \pi v_c^2}
\end{equation}
Here $A_c = \frac{ 8 q^2}{ v_c^4 \beta^2 \sigma}$. It is not possible to solve the equations arising due to Eq.~(\ref{critical}) completely for $v_c$ due to the nature of the equations. However, one could use ${\it Mathematica}$ to solve for $P_c, T_c$ and $v_c$ for  numerical values for $\beta$ and $\sigma$. The number $\frac{ P_c v_c}{ T_c}$ is an interesting number to calculate. In the Van der Waals fluids, $\frac{ P_c v_c}{ T_c}= \frac{3}{8}$ is a universal number for all fluids. It was shown that the Reissner-Nordstr\"om-AdS black hole undergo criticality in Ref.~\cite{mann} and the value for $\frac{ P_c v_c}{ T_c}$ is  $\frac{3}{8}$. In the current work, we have computed $\frac{ P_c v_c}{ T_c}$ numerically for various values of $\sigma$ and have plotted in  Fig.~$\ref{ratiovssigma}$. It is observed that the ratio $\frac{ P_c v_c}{ T_c}$ increases with $\sigma$. For $\sigma < 1$, $\frac{ P_c v_c}{ T_c} < \frac{3}{8}$ and for $\sigma >1$, $\frac{ P_c v_c}{ T_c} > \frac{3}{8}$.

\begin{figure} 
\begin{center}
\includegraphics[scale=0.9]{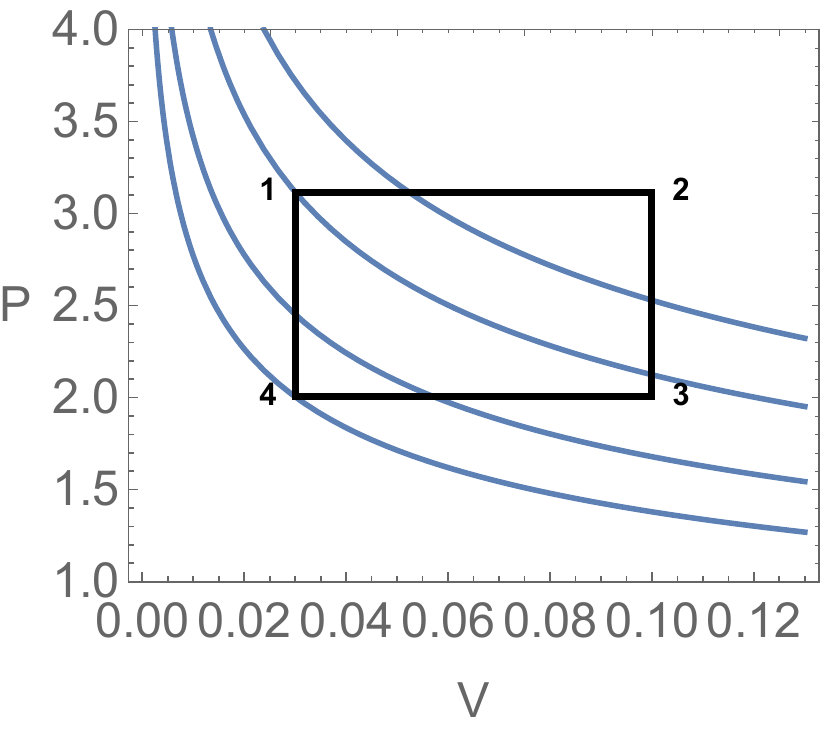}
\caption{The figure shows  P vs V for isothermals for varying temperatures. Here  $q= 1, \sigma = 0.35$, and $\beta = 0.412$}
\label{cycle}
 \end{center}
 \end{figure}

\section{Born–Infeld-type black hole as heat engine}

Black holes could be used as heat engines as demonstrated in several works mentioned in the introduction~\cite{johnson1,johnson2,johnson4,Johnson:2016pfa,yerra,sun,mo5,meng,meng2,mann2,Balart:2019uok}.  Since an equation of state is clearly defined with a pressure term and a volume, the $Pdv$ term could be utilized to produce work. In order  to create a heat engine, one need a thermodynamic cycle: we will use a thermodynamical cycle consisting of two isobaric and two isochoric paths as given in
Fig.~$\ref{cycle}$. Heat will be supplied along the path $1 \rightarrow 2$ and  heat is out from the path $ 4 \rightarrow 3$. According to Ref.~\cite{Johnson:2016pfa} the efficiency for the heat engine for such an engine is given by $\eta = 1 - \frac{M_3 - M_4}{M_2 - M_1}$.
If we use Eq.~(\ref{mass-BH-kr}), the efficiency is obtained as
\begin{equation}
\eta = \frac{8 \left(P_1-P_3\right) \sqrt{S_3} \sqrt{S_4} \left(S_3+\sqrt{S_4}
   \sqrt{S_3}+S_4\right)}{\sqrt{S_3} \sqrt{S_4} \left(8 P_1
   \left(S_3+\sqrt{S_4} \sqrt{S_3}+S_4\right)+3\right)-3 \pi  q^2  +  F(S_3,S_4)} 
\,\,\label{eff-value-kr} \,  .
\end{equation}
where 
\begin{eqnarray}
&& F(S_3,S_4) = \frac{1}{\pi  \left(\sqrt{S_3}-\sqrt{S_4}\right)} 
\left\lbrace\sqrt{S_3} \left[2 S_4^2 \beta^2
   \left(\, _{2}F_1\left(-\frac{3}{4},-\sigma;\frac{1}{4};-\frac{\pi ^2 q^2}{2 S_4^2 \beta^2
   \sigma }\right)-1    \right) \right.  \right.  
\nonumber   \\ && \left. +3 \pi ^2 q^2 \Bigg] -\sqrt{S_4}
 \Bigg.  \left[2 S_3^2  \beta^2 \left(\, _{2}F_1\left(-\frac{3}{4},-\sigma;\frac{1}{4};-\frac{\pi ^2
   q^2}{2 S_3^2 \beta^2 \sigma }\right)-1\right)+3 \pi ^2
   q^2\right]\right\rbrace
\,\,\label{F-eff-value-kr} \,  .
\end{eqnarray}
Note that when $\beta \rightarrow \infty$, $F(S_3,S_4) = 0$. Hence, $\eta \rightarrow \eta_0$ where,
\begin{equation}
\eta_0 = \frac{8 \left(P_1-P_3\right) \sqrt{S_3} \sqrt{S_4} \left(S_3+\sqrt{S_4}
   \sqrt{S_3}+S_4\right)}{\sqrt{S_3} \sqrt{S_4} \left(8 P_1
   \left(S_3+\sqrt{S_4} \sqrt{S_3}+S_4\right)+3\right)-3 \pi  q^2 } 
\,\,\label{eff-value-RN} \,  .
\end{equation}
Here $\eta_0$ is the efficiency  of the Reissner-Nordstr\"om-AdS black hole solution. Also notice that,
\\
$F(S_3,S_4) > 0$ and $\eta < \eta_0$ \, ,\, if $\sigma < 1$ ($\sigma = \frac{1}{2}$ Born-Infeld case)
\\
$F(S_3,S_4) < 0$ and $\eta > \eta_0$ \, ,\, if $\sigma > 1$ 
\,\,\,

\begin{figure}
\begin{center}
\includegraphics{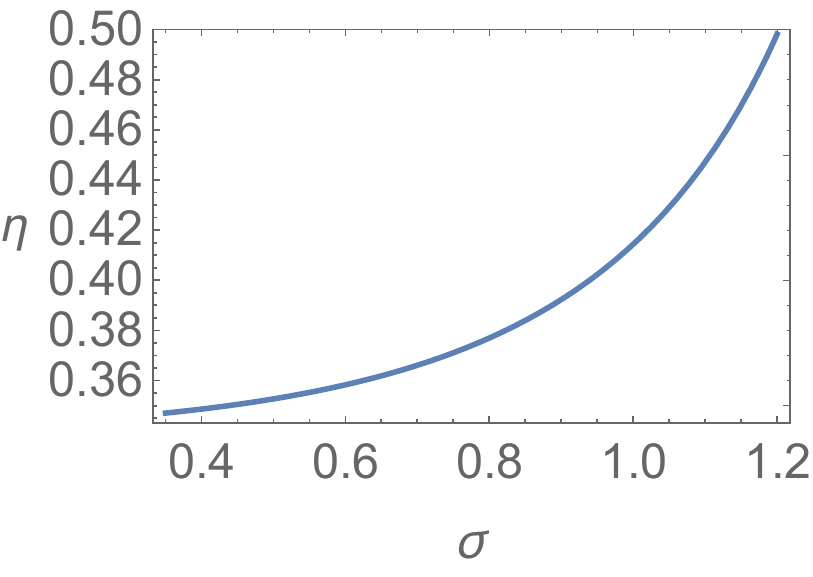}
\caption{The figure shows  $\eta$ vs $\sigma$. Here  $q= 1, \beta = 0.412, P_1 = 3.117, P_3 = 2.004, S_3 = 1.199$, and  $S_4 = 0.537$}
\label{effi}
 \end{center}
 \end{figure}

\begin{figure}
\begin{center}
\includegraphics{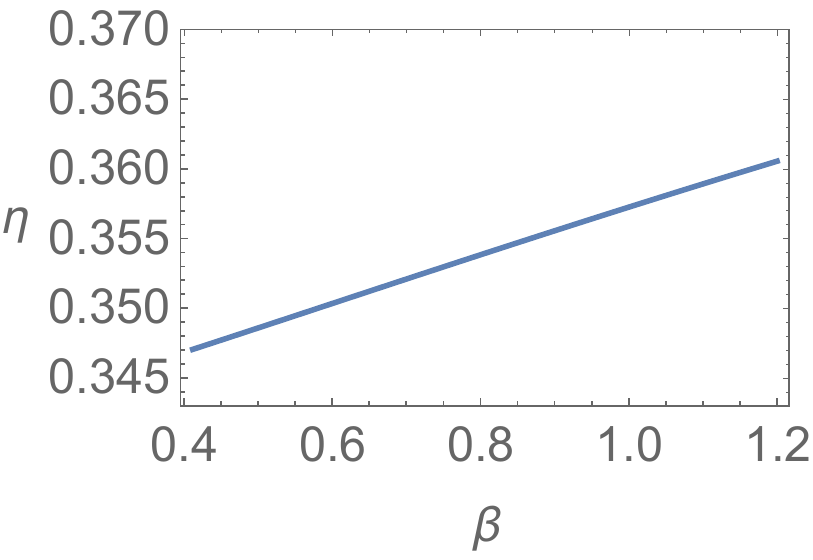}
\caption{The figure shows  $\eta$ vs $\beta$. Here  $q= 1, \sigma = 0.35, P_1 = 3.117, P_3 = 2.004, S_3 = 1.199$, and  $S_4 = 0.537$}
\label{effibeta}
 \end{center}
 \end{figure}

We have plotted $\eta$ vs $\sigma$ in Fig.~$\ref{effi}$. It is clear that when $\sigma$ increases, $\eta$ increases. We have  also plotted $\eta$ vs $\beta$ in Fig.~$\ref{effibeta}$. When $\beta$ increases, $\eta$ also increases. Hence for fixed $\beta$, large $\sigma$ yields greater $\eta$.

\section{Conclusions}
In this paper, we have studied black hole solutions in a new non-linear electrodynamics model introduced by Kruglov \cite{krug1}. There are basically two types of black holes for this new model: for small $\sigma$ the black hole behaves as the Schwarzschild-AdS black hole, and, for large $\sigma$, the black hole behaves as the Reissner-Nordstr\"om-AdS black hole.  When $\sigma = 1/2$, the black hole simplifies to the Born-Infeld black hole. 

Thermodynamics of the black hole is studied in the extended phase space where $P = - \frac{ \Lambda}{ 8 \pi}$. First law and the Smarr formula are derived: here, there is a new thermodynamical quantity introduced as a conjugate quantity to $\beta$. We have shown that the black hole demonstrate phase transitions similar to van der Waals liquid-gas phase transitions. These black holes are between small and large black holes. It is observed that the ratio $\frac{ P_c v_C}{T_c}$ is smaller than the universal values for liquids, $\frac{3}{8}$, when $\sigma < 1$. For large $\sigma >1$ the ratio is large than $\frac{3}{8}$. 

In Ref.~\cite{mann4}, it was indicated that the quantity $B$ in the case of Born-Infeld black holes has units of electric polarization per unit volume. Such quantity is called Born-Infeld polarization. In this same reference it is shown that $B$ is always positive and also noted that its physical meaning is not yet fully understood. For the model we have considered, it can be shown that $B(r_+) \geq 0$ if $\sigma \leq 1$ and $B(r_+) < 0$ if $\sigma > 1$.  It would be interesting to ask about the interpretation of $B$ when its value is negative. We leave this aspect of the theory for future work.

We have studied these black holes as heat engines. The thermodynamical cycle considered  for the heat engine is a rectangle in $P, V$ space with two isochoric and two isobaric processes. The efficiency of the heat engine, when considered the black hole as a heat engine, is studied. At this point, as a global conclusion, we can point out the relationship between the efficiency and the electric field, in particular with the values it takes with respect to the Coulomb field if we consider the same electric charge in both cases. Thus, via a numerical analysis, we can observe that by varying the parameter $\sigma$, which determines the regularity of the electrodynamic model, and keeping $\beta$ fixed (for the same electric charge), the efficiency increases together with the amplification of the electric field with respect to the Coulomb electric field. We also observe that in the region of parameter values corresponding to non-regular electric field solutions, the efficiency takes its highest values, for this family of Born-Infeld type solutions; unlike what happens in the region of regularity of the electric field, where the efficiency decreases as the electric field is increasingly screened with respect to the Coulomb electric field (the Born-Infeld case belongs to this region). In this region of regularity, we can also notice that as the value of the intercept of the curve of the electric field graph decreases, at the same time the efficiency decreases.

Additionally, let us indicate that within each of the two regions: of regularity or non-regularity, which are determined by the parameter $\sigma$ and in no case of $\beta$. The electric field can increase or decrease by varying the parameter $\beta$, with the limiting that it does not reach values that allow it to leave the region determined by $\sigma$. In summary, in each region, we can vary one or both $\sigma$ and $\beta$ parameters at the same time, and if this makes the electric field of the model increase, for example, then the efficiency value also increases.

It would be interesting to study these black holes in the context of stability due to perturbations and see how it is related to phase transitions and the $\sigma$ value.


\section*{Acknowledgments}

L.B. was supported by DIUFRO through the project DI19-0052.


\end{document}